\documentclass[preprint,12pt]{elsarticle}

\usepackage{epsfig}

\usepackage{amssymb}
\usepackage{comment}

\journal{Annals of Physics}

\begin{document}

\begin{frontmatter}



\title{Resonant Transmission in One-Dimensional Quantum Mechanics with Two Independent Point Interactions: Full Parameter Analysis}


\author{Kohkichi Konno}
\ead{kohkichi@tomakomai-ct.ac.jp}

\author{Tomoaki Nagasawa}
\ead{nagasawa@tomakomai-ct.ac.jp}

\author{Rohta Takahashi}
\ead{takahashi@tomakomai-ct.ac.jp}

\address{National Institute of Technology, Tomakomai College, 
             443 Nishikioka, Tomakomai 059-1275, Japan}

\begin{abstract}

We discuss the scattering of a quantum particle by two independent successive point interactions in one dimension.
The parameter space for two point interactions is given by $U(2)\times U(2)$, which is described by eight real parameters.
We perform an analysis of perfect resonant transmission on the whole parameter space.
By investigating the effects of the two point interactions on the scattering matrix of plane wave, 
we find the condition under which perfect resonant transmission occurs.
We also provide the physical interpretation of the resonance condition.

\end{abstract}

\begin{keyword}

one-dimensional quantum systems \sep transmission \sep resonance  


\PACS 03.65.-w \sep 03.65.Xp \sep 03.65.Db


\end{keyword}

\end{frontmatter}


\section{Introduction}
\label{sec1}


One-dimensional quantum systems with point interactions are quite non-trivial.
The point interaction in one-dimensional quantum systems has a relatively large parameter space, in comparison with those in higher dimensions.
It has been known that a point interaction in one dimension is parametrized by the group $U(2)$ \cite{U21, U22, U23}, while that in two or three dimensions is parametrized by $U(1)$. 
The parameters characterize connection conditions for a wavefunction and
its derivative. In one dimension, a variety of connection conditions leads to various intriguing physical properties such as duality \cite{duality1, duality2}, anholonomy \cite{AH1}, supersymmetry \cite{susy1, susy2, susy3, susy4, susy5}, geometric phase \cite{GP1, GP2, GP3}, and scale anomaly \cite{AH1}.

We consider the scattering of a quantum particle by point interactions in one dimension.
Several authors  \cite{SC16, SC2, SC3, SC4, SC15, SC5, SC6, SC7, SC8,
SC10, SC11, SC12, SC13, SC14} have been investigated the scattering properties by potential barriers made of the Dirac delta functions and its (higher) derivatives.
Essential properties of the scattering by a single point interaction parametrized by the $U(2)$ were investigated in \cite{GP2, AH1, Mobius, SC1}.
The authors of \cite{cheon} discussed the scattering by scale-invariant
point interactions, which are considered to be a subclass of the point
interactions.
The parameter space of a scale-invariant point interaction is given by a sphere $S^2$, which is described by two parameters.
They showed that the quantum transmission through arbitrary $N$ scale invariant point interactions exhibits random quantum dynamics.
In our previous paper \cite{KNT},  
we investigated the scattering by two independent, successive parity-invariant point interactions in one dimension.
The parameter space of a parity-invariant point interaction is given by a torus $T^2 = S^1 \times S^1$. 
Thus the parameter space of two independent parity-invariant point interactions is given by $T^2 \times T^2$, which is described by four real parameters.
Even in the reduced parameter space, it was shown that non-trivial 
resonant conditions for perfect transmission appear.
In this paper, we extend our previous work \cite{KNT} to the cases of
the scattering by two independent point interactions in one dimension
without any restriction, that is, 
on the whole parameter space $U(2)\times U(2)$.
The main purposes of this paper are to investigate the conditions 
for perfect resonant transmission on the whole parameter space  
and to provide its physical interpretation.

This paper is organized as follows.
In section \ref{sec2}, we review the scattering of plane wave by a single point interaction, and give the scattering matrix formula.
In section \ref{sec3}, we consider the scattering by two independent point interactions, derive the scattering amplitudes and the transmission probability, and investigate the conditions for the parameter space under which perfect resonant transmission occurs.
Furthermore, the physical interpretation of the perfect resonant transmission condition is discussed.
Finally, section \ref{sec4} is devoted to a summary.

\section{One-dimensional quantum systems with a point interaction}
\label{sec2}

\subsection{Connection conditions and parametrization}

In this section,
we discuss quantum mechanics in one dimension ($x$-axis) with a point interaction located at $x=\xi$.  
A point interaction is specified by a characteristic matrix $U\in U(2)$,
 and a wavefunction $\varphi(x)$ and its derivative $\varphi'(x) (=\frac{d}{dx} \varphi(x))$ are required to obey the connection conditions
 \begin{eqnarray}
	(U-I) \Phi(\xi) + iL_0 (U+I)\Phi' (\xi)=0, \label{cc}
\end{eqnarray}
where $I$ is the $2\times 2$ identity matrix. 
The parameter $L_0$ is an arbitrary nonzero constant with the dimension of length, and 
 \begin{eqnarray}
	\Phi(\xi)=\left(  \begin{array}{c}
		\varphi(\xi_+) \\
		\varphi(\xi_-) 
	\end{array}
	\right), \ \ \Phi'(\xi) = \left(
		\begin{array}{c}
			\varphi'(\xi_+) \\
			-\varphi'(\xi_-)
		\end{array}
			 \right),
\end{eqnarray}
where $\xi_{\pm}$ denotes $\xi \pm \epsilon$ with an infinitesimal positive constant $\epsilon$.
The parameter $L_0$ does not provide an additional freedom independent of the characteristic matrix $U$(see below for the details).
The probability current
\begin{eqnarray}
	j(x) = -\frac{i \hbar}{2m} (\varphi^* \varphi'- (\varphi^*)' \varphi )(x),
\end{eqnarray}
is continuous around the singular point, i.e., 
\begin{eqnarray}
	j(\xi_-) &=& j(\xi_+), 
\end{eqnarray}
under the connection conditions.

Any $2 \times 2$ unitary matrix $U \in U(2)$ can be parametrized as (see Appendix A)
\begin{eqnarray}
	U(\theta_+, \theta_-, \mu, \nu)=V^{\dagger} D V  \label{U1},
\end{eqnarray}
where
\begin{eqnarray}
	&&D=\left(
		\begin{array}{cc}
			e^{i\theta_+} & 0\\
			0 & e^{i\theta_-} 
		\end{array}
		\right), \ \ 
	V= e^{i\frac{\mu}{2}\sigma_2} e^{i \frac{\nu}{2}\sigma_3}, \label{U2} \\
	&&\mu \in [0,\pi],\ \ \nu \in [0, 2\pi),\ \ \theta_{\pm} \in [0,2\pi). \label{U3}
\end{eqnarray}
Here $\sigma_i$ ($i=1,2,3$) denotes the Pauli matrices, 
\begin{eqnarray}
	\sigma_1 = \left( 
		\begin{array}{cc}
			0 & 1 \\
			1 & 0
		\end{array}
	\right),\ \ 
	\sigma_2 = \left(
		\begin{array}{cc}
			0 & -i \\
			i & 0
		\end{array}
	\right),\ \ 
	\sigma_3 = \left(
		\begin{array}{cc}
			1 & 0 \\
			0 & -1
		\end{array}
	\right).\ \ \label{sigma}
\end{eqnarray}
The characteristic matrix $U$ can be explicitly written by 
\begin{eqnarray}
	U =\frac{1}{2}
		\left(
			\begin{array}{cc}
		(e^{i\theta_+}+e^{i\theta_-}) + \cos \mu (e^{i\theta_+}- e^{i\theta_-}) &
		e^{-i \nu}\sin \mu (e^{i\theta_+} - e^{i\theta_-}) \\
		e^{i\nu} \sin \mu (e^{i\theta_+}- e^{i \theta_-}) &
		(e^{i\theta_+}+e^{i\theta_-}) -\cos \mu (e^{i\theta_+}-e^{i\theta_-})
		\end{array}
	\right).
\end{eqnarray}
Multiplying Eq. (\ref{cc}) by $V$ from the left, we have
\begin{eqnarray}
	(D-I) V\Phi (\xi) + iL_0 (D+I ) V\Phi'(\xi) =0. \label{cc-2}
\end{eqnarray}
Equation (10) is written as
\begin{eqnarray}
	e^{i \nu} \cos \frac{\mu}{2} \varphi(\xi_+) +\sin \frac{\mu}{2}\varphi(\xi_-) + L^{(+)}  \left( e^{i\nu} \cos \frac{\mu}{2} \varphi'(\xi_+) - \sin \frac{\mu}{2} \varphi'(\xi_-)\right) &=&0 ,\nonumber \\
	&&\label{CC1} \\
	-e^{i \nu} \sin \frac{\mu}{2} \varphi(\xi_+) + \cos \frac{\mu}{2} \varphi(\xi_-)- L^{(-)}  \left( e^{i\nu} \sin \frac{\mu}{2} \varphi'(\xi_+)+ \cos \frac{\mu}{2} \varphi'(\xi_-) \right) &=&0,\nonumber \\
	&& \label{CC2}
\end{eqnarray}
 where
  \begin{eqnarray}
 	L^{(\pm)} \equiv L_0 \cot \frac{\theta_{\pm}}{2}. \label{Lpm}
 \end{eqnarray}
The parameters $\theta_+$ and $\theta_-$ appear only in the expression of $L^{(\pm)}$ in the connection conditions.
Therefore the freedom of changing the value $L_0$ can be absorbed by the corresponding change in the parameters $\theta_+$ and $\theta_-$.
Thus the parameter $L_0$ does not provide an additional freedom independent of the $U$.

We note that the characteristic matrix with interchange between $\theta_+$ and $\theta_-$ is equivalent to that with appropriate choice of $\mu$ and $\nu$, i.e., 
\begin{eqnarray}
	U(\theta_-, \theta_+ , \mu, \nu) = U(\theta_+, \theta_- , \pi- \mu, \nu+ \pi).
\end{eqnarray}	
Thus the parameter domain in Eqs. (\ref{U1}), (\ref{U2}), and (\ref{U3}) doubly covers the entire $U(2)$.
Hence, we restrict the parameter space as 
\begin{eqnarray}
	L^{(+)} \in (-\infty, \infty), \ \ L^{(-)} \in (-\infty, L^{(+)}], \ \ \mu \in [0,\pi],\ \ \nu\in [0,2\pi), \label{4p}
\end{eqnarray}
which mean $L^{(+)}-L^{(-)}\ge 0$.
Therefore, the connection conditions at a point interaction are specified  by the four parameters in Eq. (\ref{4p}).

 We provide characteristic examples for the connection conditions.
\begin{description}

\item[(i)] $L^{(+)}=L^{(-)}=L$ or $\mu=0$ or $\mu=\pi$ \\
In these cases, the connection conditions reduce to 
\begin{eqnarray}
	&&\left\{ 
		\begin{array}{c} \varphi(\xi_+) + L \varphi'(\xi_+)=0 ,\\
			 \varphi(\xi_- ) - L \varphi'(\xi_-) =0,
		\end{array}
	\right.\ \ {\rm for}\ \ L^{(+)}=L^{(-)}=L, \\
	&&\left\{ 
		\begin{array}{c} 
			\varphi(\xi_+) + L^{(+)} \varphi'(\xi_+)= 0, \\
			\varphi(\xi_-) - L^{(-)} \varphi'(\xi_-) =0,
		\end{array}
	\right.\ \ {\rm for}\ \ \mu=0,\\
	&&\left\{ 
		\begin{array}{c} 
			\varphi(\xi_+) + L^{(-)} \varphi'(\xi_+)= 0, \\
			\varphi(\xi_-) - L^{(+)} \varphi'(\xi_-) =0,
		\end{array}
	\right.\ \ {\rm for}\ \ \mu=\pi.
\end{eqnarray}
These lead to $j(\xi_+)=j(\xi_-)=0$, i.e., the probability current vanishes at $x=\xi$.
These point interactions are those of a perfect wall located at $x=\xi$, through which no probability flow is permitted.

\item[(ii)] $ \mu =\frac{\pi}{2}$ and $\nu= 0$\\
In this case, the connection conditions reduce to
\begin{eqnarray}
	&&\varphi(\xi_+) + \varphi(\xi_-) + L^{(+)} \left( \varphi'(\xi_+) -\varphi'(\xi_-) \right)=0, \\
	&& \varphi(\xi_+) - \varphi(\xi_-) + L^{(-)} \left( \varphi'(\xi_+) +\varphi'(\xi_-) \right) =0.
\end{eqnarray}
This is the parity invariant connection conditions derived in \cite{KNT}.
Furthermore, when $L^{(-)} =0$, these become
\begin{eqnarray}
	&&\varphi(\xi_+) = \varphi(\xi_-), \\
	&& \varphi'(\xi_+) - \varphi'(\xi_-) = -\frac{2}{L^{(+)}}\varphi(\xi_+). 
\end{eqnarray}
This gives a potential by the Dirac delta function.
\end{description}

\subsection{Scattering matrix}

We discuss the scattering of plane wave by the point interaction located at $x=\xi$, shown in Fig.1. 
\begin{figure}[tb]
\begin{center}
\includegraphics[width=80mm]{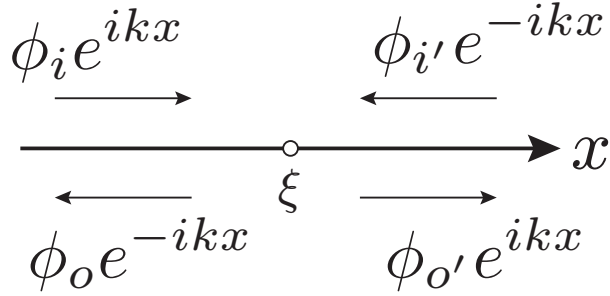}
	\caption{ One dimensional space with a point interaction, which is located at $x=\xi$.
	The input amplitudes  are $\phi_i$ and $\phi_{i'}$  and the output amplitudes are $\phi_o$ and $\phi_{o'}$.
	 }
\end{center}
\end{figure}
We assume the wave functions as 
\begin{eqnarray}
	\varphi(x)= \left\{
		\begin{array}{ll}
			\phi_i e^{ikx}+ \phi_{o}e^{-ikx} & (x< \xi) \\
			\phi_{i'} e^{-ikx}+\phi_{o'}e^{ikx} & (x > \xi)
		\end{array}
		\right. , \label{WF1}
\end{eqnarray}
where $k\ (>0)$ denotes the wave number.
The scattering matrix $S[\xi]$ is defined by 
\begin{eqnarray}
	\left(
		\begin{array}{c}
			\phi_{o'} \\
			\phi_o
		\end{array}
	\right) 
	&=& S[\xi] \left( 
		\begin{array}{c}
			\phi_{i'} \\
			\phi_i
		\end{array}
		\right), \\
	S[\xi]&=& \left(
		\begin{array}{cc}
			R_{\xi}^{(r)} & T_{\xi}^{(l)} \\
			T_{\xi}^{(r)} & R_{\xi}^{(l)}
		\end{array}
		\right). \label{SM}
\end{eqnarray}
The $R_{\xi}^{(l)}$ ($R_{\xi}^{(r)}$) and $T_{\xi}^{(l)}$ ($T_{\xi}^{(r)}$) mean the reflection and transmission amplitudes from the plane wave incoming from the left (right).
Substituting Eq. (\ref{WF1}) into 
Eqs. (\ref{CC1}) and (\ref{CC2}), we obtain the components of the scattering matrix \cite{Mobius}, explicitly as 
\begin{eqnarray}
	R_{\xi}^{(r)} 
		&=&- \frac{(1+k^2L^{(+)}L^{(-)})- ik \cos\mu (L^{(+)}- L^{(-)})}{(1+ik L^{(+)})(1+ik L^{(-)})} e^{-i2k\xi}, \label{SA1}\\
	R_{\xi}^{(l)} &=& -\frac{(1+k^2L^{(+)}L^{(-)}) +ik \cos \mu (L^{(+)}- L^{(-)})}{(1+ik L^{(+)})(1+ik L^{(-)})} e^{i2k \xi}, \label{SA2}\\
	T_{\xi}^{(r)} &=&	\frac{ik e^{i\nu} \sin\mu (L^{(+)}- L^{(-)})}{(1+ik L^{(+)})(1+ik L^{(-)})}, \label{SA3}\\ 
	T_{\xi}^{(l)} &=& \frac{ik e^{-i\nu} \sin\mu (L^{(+)}- L^{(-)})}{(1+ik L^{(+)})(1+ik L^{(-)})}. \label{SA4}
\end{eqnarray}
These satisfy the relations,
\begin{eqnarray}
	\left( R_{\xi}^{(r)}\right)^* R_{\xi}^{(r)} + \left( T_{\xi}^{(r)}\right) ^* T_{\xi}^{(r)} &=&1, \label{UnitarityOfS1} \\
	\left( R_{\xi}^{(l)}\right)^* R_{\xi}^{(l)} + \left( T_{\xi}^{(l)}\right) ^* T_{\xi}^{(l)} &=&1,  \label{UnitarityOfS2}\\
	\left( T_{\xi}^{(l)}\right)^* R_{\xi}^{(r)} + \left( R_{\xi}^{(l)}\right)^* T_{\xi}^{(r)} &=&0,\label{UnitarityOfS3}
\end{eqnarray}
which are the consequences of the unitarity  property of the S-matrix, i.e. $S^{\dagger}S=I$.
The transmission probability ($T_1$) and the reflection probability ($R_1$) are calculated as
\begin{eqnarray}
	T_1&=& | T_{\xi}^{(r)} |^2 = | T_{\xi}^{(l)}|^2, \\
	&=& \frac{k^2 \sin^2\mu (L^{(+)}-L^{(-)})^2}{(1+(kL^{(+)})^2) (1+(kL^{(-)})^2)},\\
	R_1&=& |R_{\xi}^{(r)}|^2= |R_{\xi}^{(l)}|^2 ,\\
	&=& \frac{(1+k^2 L^{(+)}L^{(-)})^2 +k^2 \cos^2 \mu (L^{(+)}-L^{(-)})^2}{(1+(kL^{(+)})^2) (1+(kL^{(-)})^2)}.
\end{eqnarray}
From these results, 
we find that the transmission and reflection probabilities are irrelevant to the parameter $\nu$ (see also \cite{Mobius}). 

The behavior of the transmission probability ($T_1$) depends on the value of $L^{(\pm)}$.
When $L^{(+)}=L^{(-)}$, the transmission possibility vanishes ($T_1=0$) at any $k$.
When $L^{(+)}=0$, 
we obtain
\begin{eqnarray}
	T_1(L^{(+)}=0) = \frac{k^2 \sin^2 \mu \left( L^{(-)} \right)^2}{1+\left( kL^{(-)} \right) ^2 }.
\end{eqnarray}	
Thus we have $T_1 =0$ at $k=0$ and $T_1 \to \sin^2 \mu$ at $k \to \infty$.
$T_1$ behaves similarly in the case of $L^{(-)}=0$.
When $L^{(+)}\to \infty$, we obtain
\begin{eqnarray}
	T_1 (L^{(+)}\to \infty)= \frac{\sin^2 \mu}{1+ \left( kL^{(-)}\right)^2 } .
\end{eqnarray}
Thus we have $T_1=\sin^2 \mu$ at $k=0$ and $T_1 =0$ at $k \to \infty$.
When $L^{(+)} \to \infty$ and $L^{(-)}=0$ (or $L^{(+)}=0$ and $L^{(-)} \to -\infty)$, 
the probability becomes constant with respect to $k$, 
\begin{eqnarray}
	T_1= \sin^2 \mu.
\end{eqnarray}
 This is because the theory is invariant under the scale transformation, since the scale parameter $L_0$ disappears in the connection conditions. 
In all other cases, the transmission probability vanishes at both $k=0$ and $k\to \infty$, and has the peak at $k=\sqrt{\frac{1}{L^{(+)}L^{(-)}}}$ for $L^{(+)}L^{(-)}>0$ or at $k=\sqrt{-\frac{1}{L^{(+)}L^{(-)}}}$ for $L^{(+)}L^{(-)}<0$ (see Fig.2).

\begin{figure}[ht]
 \begin{center}
 	\includegraphics[scale=0.8]{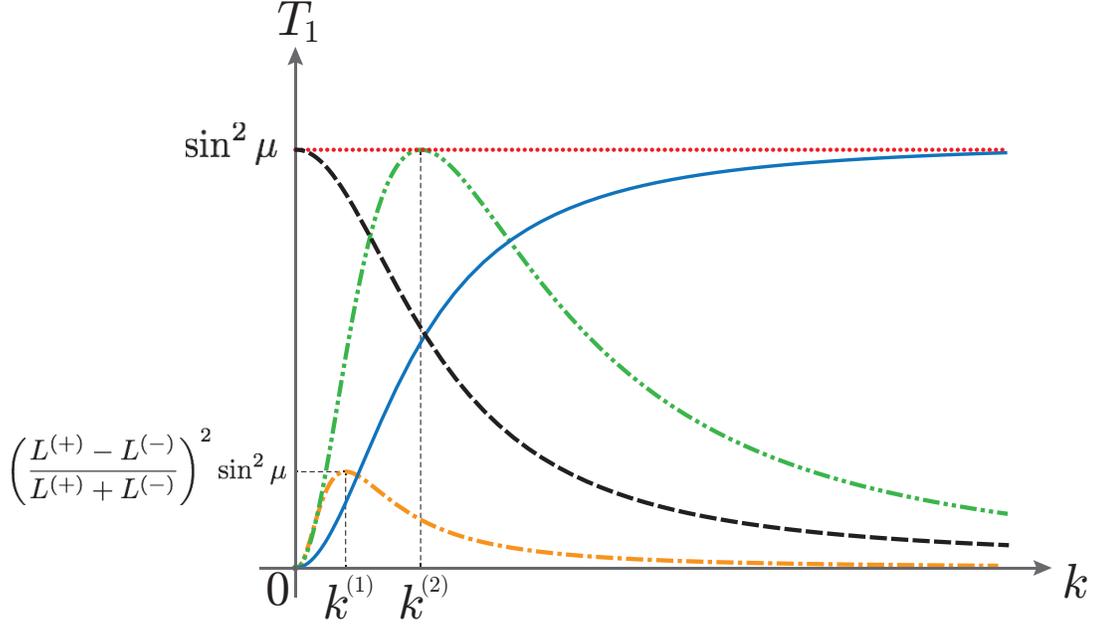} 
\end{center}
\caption{The behavior of the transmission probability is shown with respect to $k$ in the case of a single point interaction. 
The solid (blue) curve corresponds to $L^{(+)}=0$ ( or $L^{(-)}=0$), the dashed (black) curve corresponds to $L^{(+)}\to \infty$, and the dotted (red) curve corresponds to $L^{(+)}\to \infty$ and $L^{(-)}=0$ ( or $L^{(+)}=0$ and $L^{(-)}\to -\infty$).
The dashed dotted (orange) curve corresponds to $L^{(+)}L^{(-)}>0$, and the peak of the transmission probability is $T_1=\left( \frac{L^{(+)}-L^{(-)}}{L^{(+)}+L^{(-)}}\right)^2 \sin^2 \mu$ at $k=k^{(1)}=\sqrt{\frac{1}{L^{(+)}L^{(-)}}}$. 
The dashed double-dotted (green) curve corresponds to $L^{(+)}L^{(-)} <0$, and the peak of the transmission probability is $T_1=\sin^2 \mu$ at $k=k^{(2)}=\sqrt{-\frac{1}{L^{(+)}L^{(-)}}}$.
}
\end{figure}

We can also define the transfer matrix $T[\xi]$ with the notation in Fig.1, as
\begin{eqnarray}
	\left(
		\begin{array}{c}
			\phi_{o'} \\
			\phi_{i'}
		\end{array}
	\right)
	&=& T[\xi] 
	\left(
		\begin{array}{c}
			\phi_i \\
			\phi_o
		\end{array}
	\right),  \ \ \
	\left( 
	\begin{array}{c}
		\phi_i \\
		\phi_o 
	\end{array}
	\right) 
	= \left( T[\xi]\right)^{-1} 
	\left( 
		\begin{array}{c}	
			\phi_{o'} \\
			\phi_{i'}
		\end{array}
	\right).
\end{eqnarray}
When we use the scattering amplitudes, $T[\xi]$ can be written as
\begin{eqnarray}
	T[\xi] &=& \left(
		\begin{array}{cc}
			\frac{1}{\left( T_{\xi}^{(l)}\right)^*} & \frac{R_{\xi}^{(r)}}{T_{\xi}^{(r)}} \\
			-\frac{R_{\xi}^{(l)}}{T_{\xi}^{(r)}} & \frac{1}{T_{\xi}^{(r)}}
		\end{array}
		\right), \ \ \
	\left(T[\xi]\right)^{-1} = \left(
		\begin{array}{cc}
			\frac{1}{ T_{\xi}^{(l)}} & - \frac{R_{\xi}^{(r)}}{T_{\xi}^{(l)}} \\
			\frac{R_{\xi}^{(l)}}{T_{\xi}^{(l)}} & \frac{1}{\left(T_{\xi}^{(r)}\right)^*}
		\end{array}
		\right).
\end{eqnarray}
The transfer matrix is also useful to study in multiple point interactions case.

\section{One-dimensional quantum systems with two point interactions}
\label{sec3}
\subsection{Scattering amplitude and transmission probability}
\begin{figure}[ht]
\begin{center}
	\includegraphics[width=100mm]{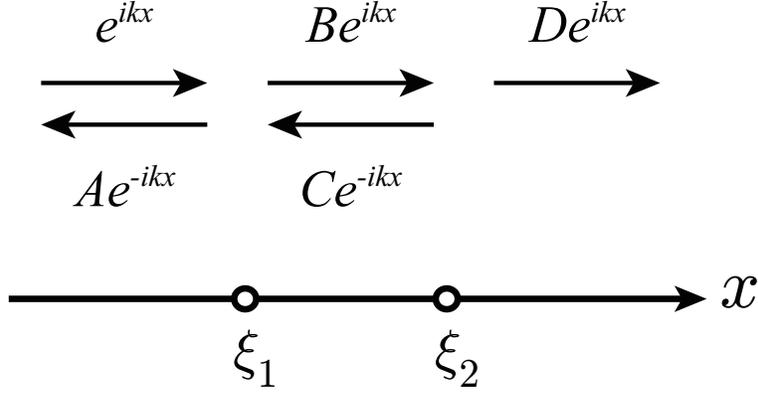}
	\caption{One dimensional space with two point interactions, which are located at $x=\xi_1$ and $x=\xi_2$. 
	The incident wave from the left-hand side is scattered at the points of $x=\xi_1$ and $x=\xi_2$.}
\end{center}
\end{figure}
In this section, we discuss the quantum mechanics in one dimension with two point interactions located at $x=\xi_1$ and $\xi_2$ $(\xi_1 < \xi_2)$.
As with the previous chapter, we describe the connection conditions at $x=\xi_i$ as four parameters, $L_i^{(\pm)}, \mu_i$, and $\nu_i$ ($i=1,2$).
Thus, there are eight parameters in this system.
The connection conditions can be explicitly written as, for $i=1,2$, 
\begin{eqnarray}
	e^{i \nu_i} \cos \frac{\mu_i}{2} \varphi(\xi_{i+}) +\sin \frac{\mu_i}{2}\varphi(\xi_{i-}) + L_i^{(+)}  \left( e^{i\nu_i} \cos \frac{\mu_i}{2} \varphi'(\xi_{i+}) - \sin \frac{\mu_i}{2} \varphi'(\xi_{i-})\right) &=&0 ,\nonumber \\
	\\
	-e^{i \nu_i} \sin \frac{\mu_i}{2} \varphi(\xi_{i+}) + \cos \frac{\mu_i}{2} \varphi(\xi_{i-})- L_i^{(-)}  \left( e^{i\nu_i} \sin \frac{\mu_i}{2} \varphi'(\xi_{i+})+ \cos \frac{\mu_i}{2} \varphi'(\xi_{i-}) \right) &=&0, \nonumber \\
\end{eqnarray}
 where $\xi_{i\pm}$ denotes 
 $\xi_i \pm \epsilon$ with an infinitesimal positive constant $\epsilon$.
The scattering matrix at $x=\xi_i$ is 
\begin{eqnarray}
	S[\xi_i] = \left(
		\begin{array}{cc}
			R_{\xi_i}^{(r)} & T_{\xi_i}^{(l)} \\
			T_{\xi_i}^{(r)} & R_{\xi_i}^{(l)}
		\end{array}
		\right) \ \ {\rm for}\  i=1,2,
\end{eqnarray} 
where
\begin{eqnarray}
	R_{\xi_i}^{(r)} 
		&=&- \frac{(1+k^2L_i^{(+)}L_i^{(-)})- ik \cos\mu_i (L_i^{(+)}- L_i^{(-)})}{(1+ik L_i^{(+)})(1+ik L_i^{(-)})} e^{-i2k\xi_i}, \label{Rir}\\
	R_{\xi_i}^{(l)} &=& -\frac{(1+k^2L_i^{(+)}L_i^{(-)}) +ik \cos \mu_i (L_i^{(+)}- L_i^{(-)})}{(1+ik L_i^{(+)})(1+ik L_i^{(-)})} e^{i2k \xi_i}, \label{Ril}\\
	T_{\xi_i}^{(r)} &=&	\frac{ik e^{i\nu_i} \sin\mu_i (L_i^{(+)}- L_i^{(-)})}{(1+ik L_i^{(+)})(1+ik L_i^{(-)})}, \label{Tir}\\ 
	T_{\xi_i}^{(l)} &=& \frac{ik e^{-i\nu_i} \sin\mu_i (L_i^{(+)}- L_i^{(-)})}{(1+ik L_i^{(+)})(1+ik L_i^{(-)})}. \label{Til}
\end{eqnarray}

We consider a plane wave incoming from the left ($x=-\infty$) with positive energy $k$.
The wave function is assumed to be
\begin{eqnarray}
	\varphi(x) = \left\{
		\begin{array}{ll}
			e^{ikx}+ A^{-ikx} & (x<\xi_1), \\
			Be^{ikx}+ C^{-ikx} & ( \xi_1 < x < \xi_2), \\
			 De^{ikx} & (\xi_2 <x),
		\end{array}
		\right.
\end{eqnarray}
where $A,B,C,$ and $D (\in \mathbb{C})$ are constants (see Fig.3).
In this setup, the scattering matrices satisfy the following relations 
\begin{eqnarray}
	\left( \begin{array}{c}	B \\ A\end{array}\right)
	=S[\xi_1] \left( \begin{array}{c} C \\ 1 \end{array}\right) ,\ \ 
	\left( \begin{array}{c}	D \\ C\end{array}\right)
	=S[\xi_2] \left( \begin{array}{c} 0 \\ B \end{array}\right). \label{ScatteringMatrices}
\end{eqnarray}
The solutions of Eqs. (\ref{ScatteringMatrices}) are
\begin{eqnarray}
	A&=& \frac{ R_{\xi_1}^{(l)} \left( 1- R_{\xi_1}^{(r)} R_{\xi_2}^{(l)} \right)+ T_{\xi_1}^{(l)} T_{\xi_1}^{(r)} R_{\xi_2}^{(l)}}{1-R_{\xi_1}^{(r)}R_{\xi_2}^{(l)}}, \label{A} \\
	B&=& \frac{T_{\xi_1}^{(l)}}{1-R_{\xi_1}^{(r)}R_{\xi_2}^{(l)}}, \\
	C&=& \frac{T_{\xi_1}^{(l)} R_{\xi_2}^{(l)}} {1-R_{\xi_1}^{(r)}R_{\xi_2}^{(l)}} ,\\
	D&=& \frac{T_{\xi_1}^{(l)} T_{\xi_2}^{(l)}}{1-R_{\xi_1}^{(r)}R_{\xi_2}^{(l)}}. \label{D}
\end{eqnarray}
These satisfy
\begin{eqnarray}
	1-|A|^2 = |B|^2-|C|^2 =|D|^2, \label{3-17}
\end{eqnarray}
because the unitarity of the scattering matrices holds.

Substituting Eqs. (\ref{Rir}), (\ref{Ril}), (\ref{Tir}), and (\ref{Til}) into Eq. (\ref{D}), we obtain the transmission amplitude $D$ as
\begin{eqnarray}
	D&=& 
		\frac{ -k^2 e^{-i (\nu_1+\nu_2)} \sin \mu_1 \sin \mu_2 \left( L_1^{(+)} - L_1^{(-)}\right) \left( L_2^{(+)} - L_2^{(-)}\right) }
		{\Delta },
\end{eqnarray}
where
\begin{eqnarray}
	\Delta &=& \left( 1+ ik L_1^{(+)}\right) \left( 1+ ik L_1^{(-)}\right)\left( 1+ ik L_2^{(+)}\right)\left( 1+ ik L_2^{(- )}\right) \nonumber \\
	&& -\left\{ (1+k^2 L_1^{(+)} L_1^{(-)} ) - ik \cos \mu_1 (L_1^{(+)}-L_1^{(-)}) \right\} \nonumber \\
	&& \times 
	\left\{ (1+k^2 L_2^{(+)} L_2^{(-)} ) + ik \cos \mu_2 (L_2^{(+)}-L_2^{(-)}) \right\}
	e^{i 2k (\xi_2-\xi_1)}.
\end{eqnarray}
The transmission amplitude $D$ vanishes when $L_1^{(+)}=L_1^{(-)}$ or $L_2^{(+)}=L_2^{(-)}$ or $\mu_1= 0$ or $\mu_1=\pi$ or  $\mu_2=0$ or $\mu_2=\pi$.
In each case, the point interaction behaves like a perfect wall, which means that probability current vanishes at the point.

Consequently, 
the transmission probability $T_2 \ ( = |D|^2)$ becomes
\begin{eqnarray}
	T_2= \frac{ k^4 \sin^2\mu_1 \sin^2 \mu_2 \left( L_1^{(+)}- L_1^{(-)}\right)^2 \left( L_2^{(+)}- L_2^{(-)}\right)^2} {|\Delta |^2 }. 
\end{eqnarray}
We note that the transmission probability is irrelevant to the parameters $\nu_1, \nu_2$.
%
%
%
\subsection{Perfect transmission}
\subsubsection{Perfect transmission condition and its physical interpretation}

The reflection amplitude $A$ can be rewritten as
\begin{eqnarray}
	A&=& \frac{\left( R_{\xi_2}^{(l)} - \left( R_{\xi_1}^{(r)}\right) ^* \right) T_{\xi_1}^{(l)}}{ \left( T_{\xi_1}^{(r)}\right)^* \left( 1-R_{\xi_1}^{(r)}R_{\xi_2}^{(l)} \right)},
\end{eqnarray}
when we use  the unitarity property of the scattering matrices $S[\xi_1]$ and $S[\xi_2]$.
Thus, the  amplitude $A$ vanishes when 
\footnote{Equation (\ref{PT}) can also  be  derived by the transfer matrix approach. 
When we consider the   matrix $T = \left( \begin{array}{cc} T_{11} & T_{12} \\ T_{21} & T_{22} \end{array} \right) =\left( T[\xi_1] \right)^{-1} \left( T[\xi_2] \right)^{-1}$,
Eq. (\ref{PT}) is derived from $T_{21}=0$.}
\begin{eqnarray}
	 R_{\xi_2}^{(l)} = \left( R_{\xi_1}^{(r)} \right)^* \label{PT}.
\end{eqnarray}
This is the condition in which perfect transmission occurs due to resonance.
%

%

To clarify the physical interpretation of the resonance condition of Eq. (\ref{PT}), we consider the outgoing plane wave $De^{ikx}$ to be the interference of an infinite number of waves with various trajectories.
In this viewpoint, $De^{ikx}$ can be written as
\begin{eqnarray}
	De^{ikx}= T_{\xi_1}^{(l)}T_{\xi_2}^{(l)} \left( 1+ \left( R_{\xi_1}^{(r)} R_{\xi_2}^{(l)}\right)  +\left( R_{\xi_1}^{(r)} R_{\xi_2}^{(l)}\right)^2+\left( R_{\xi_1}^{(r)} R_{\xi_2}^{(l)}\right)^3+ \cdots \right) e^{ikx}. \label{Dseries}
\end{eqnarray}
The trajectories corresponding to each term in Eq. (\ref{Dseries}) are shown in Fig.4.
\begin{figure}[tb]
 \begin{center}
 	\includegraphics[scale=0.5]{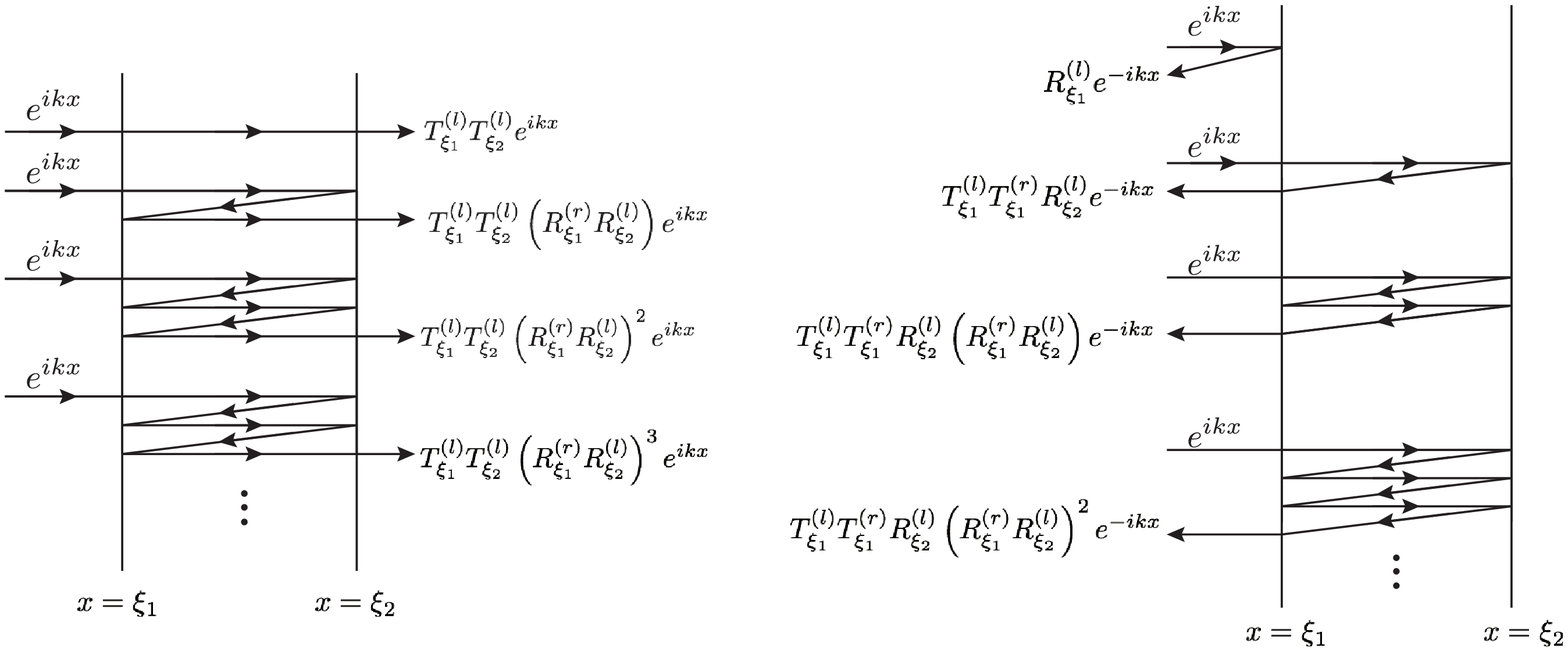} 
\end{center}
\caption{In the left figure, 
some trajectories of the transmitted plane wave are shown.
The first trajectory corresponds to the first term in Eq. (\ref{Dseries}), the second trajectory corresponds to the second term in Eq. (\ref{Dseries}), and so on.
In the right figure, some trajectories of the reflected plane wave are shown. 
The first trajectory corresponds to the first term in Eq. (\ref{61}), the sum of the second and following trajectories corresponds to the second term in Eq. (\ref{61}). 
}
\end{figure}

Since the term $ R_{\xi_1}^{(r)} R_{\xi_2}^{(l)}$, in general, takes a complex value,
$De^{ikx}$ can be interpreted  as a superposition of an infinite number of the waves with different phases.
The resonance condition of Eq. (\ref{PT}) leads to 
\begin{eqnarray}
	R_{\xi_1}^{(r)} R_{\xi_2}^{(l)} = |R_{\xi_1}^{(r)}|^2.
\end{eqnarray}
Thus, Eq. (\ref{Dseries}) becomes
\begin{eqnarray}
	De^{ikx}= T_{\xi_1}^{(l)}T_{\xi_2}^{(l)} \left( 1+  | R_{\xi_1}^{(r)} |^2  +| R_{\xi_1}^{(r)} |^4  +| R_{\xi_1}^{(r)} |^6 +\cdots \right) e^{ikx},  \label{Dseries2}
\end{eqnarray}
under the resonance condition. 
There are no phase differences among each term in  Eq. (\ref{Dseries2}).
When the resonance condition is satisfied, the outgoing plane wave can be expressed as a superposition of an infinite number of the waves with an aligned phase.

In the same manner, when the resonance condition of Eq. (\ref{PT}) holds, $Be^{ikx}$, $Ce^{-ikx}$ can also be  expressed 
 as a superposition of an infinite number of the waves with an  aligned phase, 
\begin{eqnarray}
 Be^{ikx} &=& T_{\xi_1}^{(l)}\left( 1+  | R_{\xi_1}^{(r)} |^2  +| R_{\xi_1}^{(r)} |^4  +| R_{\xi_1}^{(r)} |^6 +\cdots \right) e^{ikx} , \\
 Ce^{-ikx} &=& T_{\xi_1}^{(l)} R_{\xi_2}^{(l)} \left( 1+  | R_{\xi_1}^{(r)} |^2  +| R_{\xi_1}^{(r)} |^4  +| R_{\xi_1}^{(r)} |^6 +\cdots \right) e^{-ikx}.
\end{eqnarray}
Furthermore, the reflected plane wave $Ae^{-ikx}$ can be rewritten as
\begin{eqnarray}
	Ae^{-ikx}= R_{\xi_1} ^{(l)} e^{-ikx} + \frac{ T_{\xi_1}^{(l)}T_{\xi_1}^{(r)} R_{\xi_2}^{(l)}}{1-R_{\xi_1}^{(r)} R_{\xi_2}^{(l)}}e^{-ikx} . \label{61} 
\end{eqnarray}
When the resonance condition is satisfied, the second term can also be expressed as a superposition of an infinite number of the waves with an aligned phase,
\begin{eqnarray}
	 \frac{ T_{\xi_1}^{(l)}T_{\xi_1}^{(r)} R_{\xi_2}^{(l)}}{1-R_{\xi_1}^{(r)} R_{\xi_2}^{(l)}}e^{-ikx}
	 &=& T_{\xi_1}^{(l)}T_{\xi_1}^{(r)} R_{\xi_2}^{(l)} \left( 1+  | R_{\xi_1}^{(r)} |^2  +| R_{\xi_1}^{(r)} |^4  +| R_{\xi_1}^{(r)} |^6 +\cdots \right)e^{-ikx} . \nonumber \\
	 &&
\end{eqnarray}
In addition, when we use the resonance condition, the second term in Eq. (\ref{61}) can be written as $-R_{\xi_1} ^{(l)} e^{-ikx}$.
Thus, we find that the cancellation of the reflected plane wave occurs between the first term and the second term in Eq. (\ref{61}).

\subsubsection{Explicit expressions for perfect transmission}
When we use Eqs. (\ref{Rir}) and (\ref{Ril}), 
the perfect transmission condition of Eq. (\ref{PT}) can be written explicitly as
\begin{eqnarray}
	e^{2ik (\xi_2-\xi_1)} 
	&=& h(k,L_1^{(\pm)}, L_2^{(\pm)}, \mu_1,\mu_2),  \label{PTcondition}
\end{eqnarray}
where
\begin{eqnarray}
	&&h(k,L_1^{(\pm)}, L_2^{(\pm)}, \mu_1,\mu_2) \nonumber \\
	&&=
	\frac{ ( 1+ ikL_2^{(+)}) (1+ ikL_2^{(-)}) \left\{
		(1+k^2 L_1^{(+)} L_1^{(-)} ) + ik \cos \mu_1 ( L_1^{(+)}- L_1^{(-)})\right\}}
		{ (1-ik L_1^{(+)}) (1-ikL_1^{(-)}) \left\{
			(1+k^2 L_2^{(+)} L_2^{(-)}) +ik \cos \mu_2 (L_2^{(+)} -L_2^{(-)}) \right\}
		}.  \nonumber \\
		&&
\end{eqnarray}
The condition for the existence of a solution for Eq. (\ref{PTcondition}) is 
\begin{eqnarray}
	|h|^2=1. \label{g2}
\end{eqnarray}
This condition is expressed as 
\begin{eqnarray}
	\alpha k^4 + \beta k^2 + \gamma =0, \label{alpha}
\end{eqnarray}
where
\begin{eqnarray}
	\alpha &=&
		\sin^2 \mu_2 \left( L_2^{(+)} -L_2^{(-)} \right)^2 \left( L_1^{(+)} L_1^{(-)}\right)^2- \sin^2 \mu_1 \left( L_1^{(+)} - L_1^{(-)} \right)^2 \left( L_2^{(+)} L_2^{(-)} \right)^2, \nonumber \\
		&& \\
	\beta &=& \sin^2 \mu_2 \left( L_2^{(+)} -L_2^{(-)} \right)^2 \left( \left( L_1^{(+)}\right)^2 + \left( L_1^{(-)}\right)^2 \right) \nonumber  \\
	 &&  - \sin^2 \mu_1 \left( L_1^{(+)} -L_1^{(-)} \right)^2 \left( \left( L_2^{(+)}\right)^2 + \left( L_2^{(-)}\right)^2 \right), \\
	 \gamma &=& \sin^2 \mu_2 \left( L_2^{(+)} -L_2^{(-)} \right)^2 - \sin^2 \mu_1 \left( L_1^{(+)} - L_1^{(-)} \right)^2 .
\end{eqnarray}
When all of the coefficients in Eq. (\ref{alpha}) vanish, i.e., 
\begin{eqnarray}
	\alpha=\beta= \gamma=0, \label{abg}
\end{eqnarray}
Eq. (\ref{g2}) is identically satisfied, independent of the value of $k$.
With the definition $L_{i}^{(+)}- L_{i}^{(-)} \ge 0$ ($i=1,2$), 
we have the solutions for Eq. (\ref{abg}) as 
\begin{description}
	\item[(I)] \begin{eqnarray}
			 \sin^2 \mu_2 = \sin^2\mu_1,\ \ L_2^{(+)}=L_1^{(+)}, \ \ L_2^{(-)}=L_1^{(-)},   \label{I}
		\end{eqnarray}
	\item[(II)] 
		\begin{eqnarray}
			\sin^2 \mu_2 = \sin^2\mu_1,\ \ L_2^{(+)}= -L_1^{(-)},\ \  L_2^{(-)}= -L_1^{(+)},  \label{II}
		\end{eqnarray}
	\item[(III)] 
		\begin{eqnarray}
			&& \left( L_1^{(+)}+L_1^{(-)}\right)^2 \sin^2 \mu_2 = \left( L_1^{(+)}-L_1^{(-)}\right)^2 \sin^2 \mu_1,\ \  {\rm and} \nonumber \\
			&&\ {\rm (III-i)} \ L_2^{(+)}= L_1^{(-)}, L_2^{(-)}=-L_1^{(+)}\ \ ({\rm for}\ \ L_1^{(+)}+L_1^{(-)}>0), \nonumber \\
			&& \\
			&&\ {\rm (III-ii)} \ L_2^{(+)}=-L_1^{(+)}, L_2^{(-)}= L_1^{(-)}\ \ ({\rm for}\ \ L_1^{(+)}+ L_1^{(-)}<0), \nonumber \\ &&\\
			&&\ {\rm (III-iii)}\  L_2^{(+)}= L_1^{(+)}, L_2^{(-)}= -L_1^{(-)}\ \ ({\rm for}\ \ L_1^{(+)}+L_1^{(-)}>0),\nonumber \\ && \\
			&&\ {\rm (III-iv)} \  L_2^{(+)}= - L_1^{(-)}, L_2^{(-)}= L_1^{(+)} \ \ ({\rm for}\ \ L_1^{(+)}+L_1^{(-)}<0).\nonumber \\ &&
		\end{eqnarray}
\end{description}
When the above conditions hold, perfect transmission occurs.
The cases (I) and (II) correspond to the solutions for $L_1^{(+)}L_1^{(-)}= L_2^{(+)}L_2^{(-)}$, 
while the case (III)  corresponds to that for $L_1^{(+)}L_1^{(-)}=- L_2^{(+)}L_2^{(-)}$.

In the case (I) or (II), if $L_1^{(+)}L_1^{(-)}  (= L_2^{(+)}L_2^{(-)}) <0 $ and $\mu_1=\mu_2=\frac{\pi}{2}$, then the perfect transmission occurs at 
\begin{eqnarray}
 k= \sqrt{-\frac{1}{L_1^{(+)}L_1^{(-)}}}. \label{k}
 \end{eqnarray}
This is the case in which perfect transmission occurs at each point interaction.
Besides this, we can find an infinite number of perfect transmission peaks due to resonance. We investigate the details for each case below.\\

The case (I):\\
Since the parameter space of $\mu_i$ ($i=1,2$) is 
$ [0,\pi] $, the relation between $\mu_1$ and $\mu_2$, is divided into two cases, 
\begin{eqnarray}
	\cos \mu_1 &=& \cos \mu_2, \label{3-39} \\
	\cos \mu_1 &=& -\cos \mu_2. \label{3-40}
\end{eqnarray}
First, we consider the case of Eq. (\ref{3-39}), i.e., $\mu_1=\mu_2$. 
Substituting $L_2^{(+)}= L_1^{(+)}, L_2^{(-)}= L_1^{(-)}$, and $\cos \mu_1= \cos \mu_2$ into Eq. (\ref{PTcondition}), 
we have
\begin{eqnarray}
 e^{i 2k(\xi_2-\xi_1)} = \frac{ (1+ikL_1^{(+)}) (1+ik L_1^{(-)})}{(1-ik L_1^{(+)}) (1-ik L_1^{(-)})}.
\end{eqnarray}
This can be written as 
\begin{eqnarray}
	\tan k(\xi_2-\xi_1) 
	=f_{({\rm I})} (k),  \label{PTC-1}
\end{eqnarray}
where 
\begin{eqnarray}
	f_{({\rm I})}(k) =\frac{k (L_1^{(+)} +L_1^{(-)})}{1- k^2 L_1^{(+)} L_1^{(-)}}. \label{3-41}
\end{eqnarray}
This is the equation to determine the wave number $k$ at which perfect transmission occurs.
We note that the condition of Eq. (\ref{PTC-1}) does not contain  the parameters $\mu_1$ and $ \mu_2$.
Thus,
the value of $k$ for perfect transmission is independent of $\mu_1$ and $\mu_2$.
A representative example in this case is shown in Fig.5.
We plot the curves of the functions on each side in Eq. (\ref{PTC-1}) in the left figure. 
At the points of intersection of the solid curves and the dashed curves, perfect transmission occurs. 
Hence, we can find an infinite number of solutions for perfect transmission.
We also show the transmission probability for several values of the parameters $\mu_1$ in the right figure.
It is shown that perfect transmission occurs under the resonance condition of Eq. (\ref{PTC-1}),
and the parameter $\mu_1$ does not change the values of the $k$ at the peak, but change the peak width.
Note that  there is an extra peak at which Eq. (\ref{k}) holds, only if  $\mu_1=\mu_2=\frac{\pi}{2}$ and $L_1^{(+)}L_1^{(-)}<0$.
\begin{figure}[ht]
\begin{center}
　\includegraphics[width=160mm]{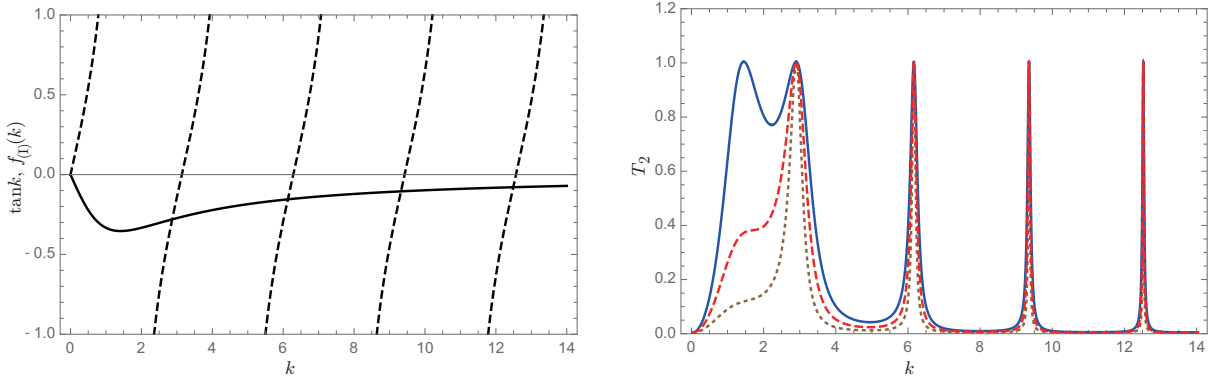}
	\caption{
	In the left figure, the solid curves correspond to the right hand side of Eq. (\ref{PTC-1}), and the dashed curves correspond to the left hand side of Eq. (\ref{PTC-1}). 
	The horizontal axis denotes the wave number $k$.
	In the right figure, the transmission probability $T_2$ is shown as a function of $k$.
	Here we adopt $L_1^{(+)}=L_2^{(+)}=0.5$, $L_1^{(-)}=L_2^{(-)}=-1$, and $\xi_2-\xi_1=1$.
	At the point of intersection of the solid curves and the dashed curves in the left figure, perfect transmission occurs.
	The solid (blue), dashed (red), and dotted (brown) curves in the right figure correspond to the $\mu_1 (=\mu_2) =\frac{\pi}{2}$, $\mu_1 (=\mu_2) =\frac{\pi}{3}$, and $\mu_1 (=\mu_2) =\frac{\pi}{4}$, respectively.
	There is another peak at $k=\sqrt{-\frac{1}{L_1^{(+)}L_1^{(-)}}}$ in the blue curve.
	 }
\end{center}
\end{figure}

Next, we consider the case of $\cos \mu_1= -\cos \mu_2$.
The resonance condition is
\begin{eqnarray}
	 \tan k (\xi_2-\xi_1) 
	 = g_{({\rm I})} (k), \label{PTC-2}
\end{eqnarray}
where
\begin{eqnarray}
 g_{({\rm I})}(k)  = 
 \frac{ k \left\{ (L_1^{(+)} + L_1^{(-)})(1+k^2 L_1^{(+)} L_1^{(-)}) +\cos \mu_1 ( L_1^{(+)} -L_1^{(-)}) (1-k^2 L_1^{(+)} L_1^{(-)})\right\}}
 	{ (1-k^4 (L_1^{(+)})^2 (L_1^{(-)})^2 ) - k^2 \cos \mu_1 (L_1^{(+)} +L_1^{(-)}) ( L_1^{(+)}- L_1^{(-)})
	}. \nonumber \\
\end{eqnarray}	
A representative example in this case is shown in Fig.6.
Similarly, we can find an infinite number of solutions for perfect transmission.
\begin{figure}[ht]
\begin{center}
　\includegraphics[width=160mm]{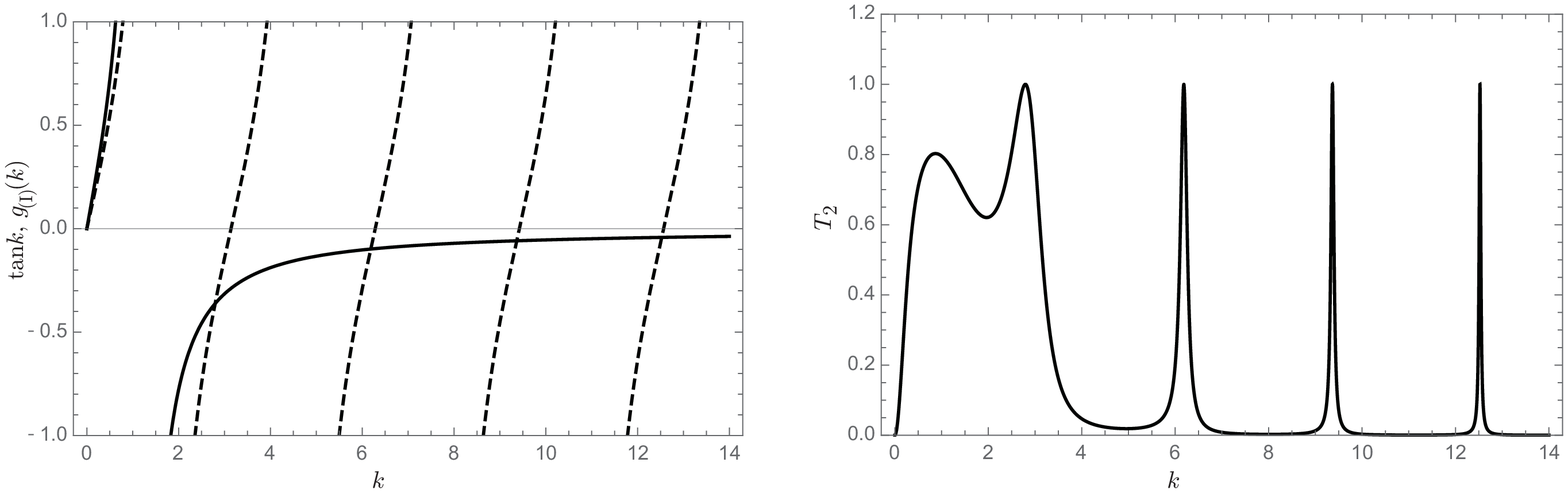}
	\caption{ 
	In the left figure, the solid curves correspond to the right hand side of Eq. (\ref{PTC-2}), and the dashed curves correspond to the left hand side of Eq. (\ref{PTC-2}). 
	The horizontal axis denotes the wave number $k$.
	In the right figure, the transmission probability $T_2$ is shown as a function of $k$.
	Here we adopt $L_1^{(+)}=1$, $L_1^{(-)}=-0.5$, $L_2^{(+)}= 1$, $L_2^{(-)}= -0.5$, $\mu_1=\frac{\pi}{3}$, $\mu_2=\frac{2\pi}{3}$, and $\xi_2-\xi_1=1$.
	At the point of intersection of the solid curves and the dashed curves in the left figure, perfect transmission occurs.
}
\end{center}
\end{figure}
\\

The case (II): \\
We consider the two cases in Eqs. (\ref{3-39}) and (\ref{3-40}) again. 
When $\cos \mu_1= \cos \mu_2$, the resonance condition becomes
\begin{eqnarray}
	\sin k(\xi_2-\xi_1) =0.
\end{eqnarray}
This leads to the solutions
\begin{eqnarray}
	k= \frac{n \pi}{\xi_2-\xi_1}\ \ (n=1,2,3,\cdots),
\end{eqnarray}
for perfect transmission.
When $\cos \mu_1= -\cos \mu_2$, the resonance condition is
\begin{eqnarray}
	\tan k(\xi_2-\xi_1) = f_{({\rm II})}(k), \label{PTC-3}
\end{eqnarray}
where
\begin{eqnarray}
	f_{({\rm II})}(k)
	=
	\frac{ k\cos \mu_1 (L_1^{(+)} - L_1^{(-)})}{1+ k^2 L_1^{(+)} L_1^{(-)}}.
\end{eqnarray}
We can find an infinite number of solutions for perfect transmission by solving Eq. (\ref{PTC-3}).
A representative example in this case is shown in Fig.7.
\begin{figure}[ht]
\begin{center}
　\includegraphics[width=160mm]{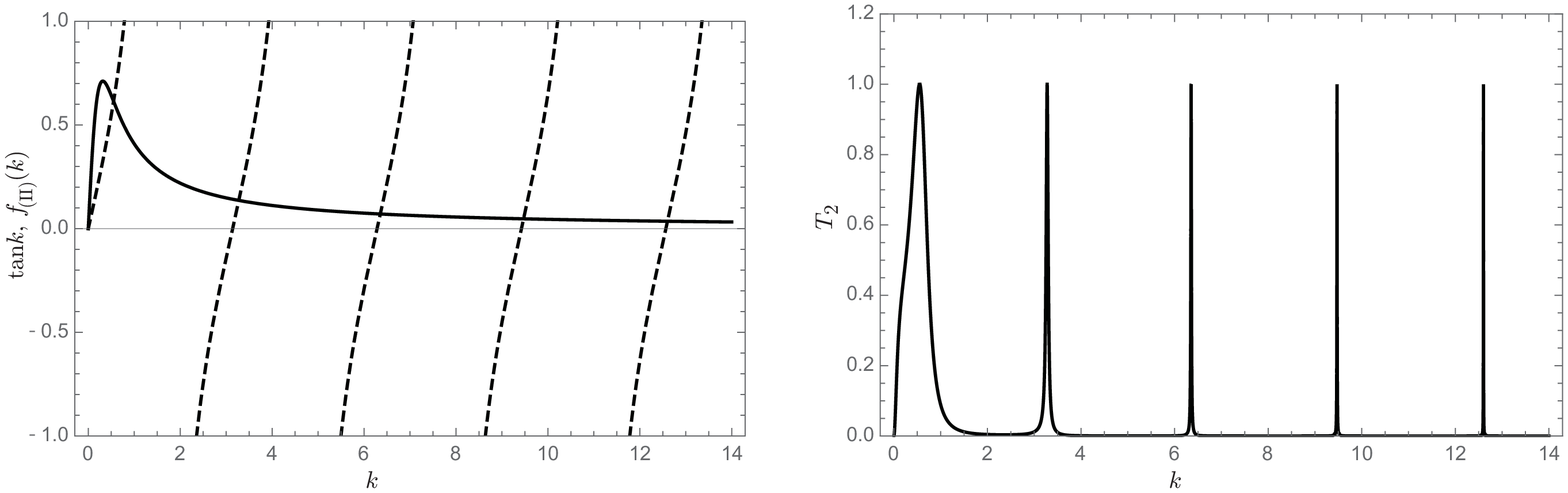}
	\caption{
	In the left figure, the solid curves correspond to the right hand side of Eq. (\ref{PTC-3}), and the dashed curves correspond to the left hand side of Eq. (\ref{PTC-3}). 
	The horizontal axis denotes the wave number $k$.	
	In the right figure, the transmission probability $T_2$ is shown as a function of $k$.
	Here we adopt $L_1^{(+)}=10$, $L_1^{(-)}=1$, $L_2^{(+)}= -1$, $L_2^{(-)}= -10$, $\mu_1=\frac{\pi}{3}$, $\mu_2=\frac{2\pi}{3}$, and $\xi_2-\xi_1=1$.
	At the point of intersection of the solid curves and the dashed curves in the left figure, perfect transmission occurs.
	 }
\end{center}
\end{figure}
\\

The case (III): \\
First, we consider the case (III-i). When
\begin{eqnarray} 
&& \left( L_1^{(+)}+L_1^{(-)}\right)^2 \sin^2 \mu_2 = \left(L_1^{(+)}-L_1^{(-)} \right)^2\sin^2 \mu_1, \label{3-51} \\
&& L_2^{(+)}=L_1^{(-)} , L_2^{(-)}=-L_1^{(+)}, \label{3-52}
\end{eqnarray}
the resonance condition becomes 
\begin{eqnarray}
	\tan k(\xi_2-\xi_1) = f_{({\rm III})}(k), \label{PTC-4}
\end{eqnarray}
where
\begin{eqnarray}
	f_{({\rm III})}(k) =
	\frac{
		kL_1^{(-)} \left( (L_1^{(+)} - L_1^{(-)}) \cos \mu_1 + (L_1^{(+)}+L_1^{(-)}) \cos \mu_2 \right) - 2k L_1^{(+)} L_1^{(-)}}
	{
	\left( (L_1^{(+)} - L_1^{(-)} ) \cos \mu_1 + (L_1^{(+)} +L_1^{(-)} ) \cos \mu_2 \right) + 2k^2 L_1^{(+)}\left( L_1^{(-)}\right)^2 
	}, \nonumber \\ && \label{fIII}
\end{eqnarray}
with 
\begin{eqnarray}
 \left( L_1^{(+)}+L_1^{(-)} \right)^2 \sin^2 \mu_2 = \left(  L_1^{(+)}- L_1^{(-)}\right)^2 \sin^2 \mu_1. \label{AA}
\end{eqnarray}
We can find an infinite number of solutions for perfect transmission by solving Eq. (\ref{PTC-4}).
A representative example is shown in Fig.8.
\begin{figure}[ht]
\begin{center}
	\includegraphics[width=160mm]{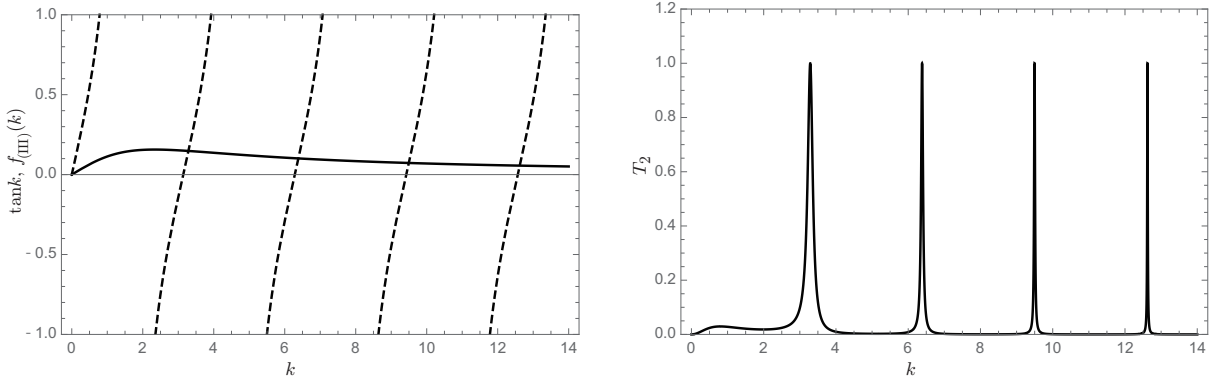}
	\caption{
	In the left figure, the solid curves correspond to the right hand side of Eq. (\ref{PTC-4}), and the dashed curves correspond to the left hand side of Eq. (\ref{PTC-4}). 
	The horizontal axis denotes the wave number $k$.
	In the right figure, the transmission probability $T_2$ is shown as a function of $k$.
	Here we adopt $L_1^{(+)}=\frac{1}{2}(1+\sqrt{3})$, $L_1^{(-)}=\frac{1}{2}(1-\sqrt{3})$, $L_2^{(+)}= \frac{1}{2}(1-\sqrt{3})$, $L_2^{(-)}= -\frac{1}{2}(1+\sqrt{3})$, $\mu_1=\frac{\pi}{6}$, $\mu_2=\frac{\pi}{3}$, and $\xi_2-\xi_1=1$, which are satisfied in Eqs. (\ref{3-51}) and (\ref{3-52}).
	At the point of intersection of the solid curves and the dashed curves in the left figure, perfect transmission occurs.
	}
\end{center}
\end{figure}
The resonance conditions for the cases (III-ii), (III-iii), and (III-iv) are provided by 
\begin{eqnarray}
	\tan k(\xi_2-\xi_1) = f_{({\rm III-j})} (k) \ \ \ ({\rm j}={\rm ii}, {\rm iii}, {\rm iv}),
\end{eqnarray}
where
\begin{eqnarray}
	f_{({\rm III-ii})}(k) &=&
	\frac{
		kL_1^{(-)} \left( (L_1^{(+)} - L_1^{(-)}) \cos \mu_1 - (L_1^{(+)}+L_1^{(-)}) \cos \mu_2 \right) - 2k L_1^{(+)} L_1^{(-)}}
	{
	\left( (L_1^{(+)} - L_1^{(-)} ) \cos \mu_1 - (L_1^{(+)} +L_1^{(-)} ) \cos \mu_2 \right) + 2k^2 L_1^{(+)}\left( L_1^{(-)}\right)^2 
	}, \label{fIII-ii} \nonumber \\
	&&\\
	f_{({\rm III-iii})}(k) &=&
	\frac{
		kL_1^{(+)} \left( (L_1^{(+)} - L_1^{(-)}) \cos \mu_1 + (L_1^{(+)}+L_1^{(-)}) \cos \mu_2 \right) - 2k L_1^{(+)} L_1^{(-)}}
	{
	\left( (L_1^{(+)} - L_1^{(-)} ) \cos \mu_1 + (L_1^{(+)} +L_1^{(-)} ) \cos \mu_2 \right) + 2k^2 \left( L_1^{(+)}\right)^2  L_1^{(-)} 
	}, \label{fIII-iii} \nonumber \\
	&&\\
	f_{({\rm III-iv})}(k) &=&
	\frac{
		kL_1^{(+)} \left( (L_1^{(+)} - L_1^{(-)}) \cos \mu_1 - (L_1^{(+)}+L_1^{(-)}) \cos \mu_2 \right) - 2k L_1^{(+)} L_1^{(-)}}
	{
	\left( (L_1^{(+)} - L_1^{(-)} ) \cos \mu_1 - (L_1^{(+)} +L_1^{(-)} ) \cos \mu_2 \right) + 2k^2 \left( L_1^{(+)}\right)^2  L_1^{(-)} 
	}. \label{fIII-iv} \nonumber \\
	&&
\end{eqnarray}
In each case, we can similarly find an infinite number of solutions for perfect transmission.
\\

Finally, it should be noticed that even if Eq. (\ref{abg}) does not hold, the positive solution $k$ satisfying the condition of Eq. (\ref{alpha}) may exist when the solution of Eq. (\ref{alpha}) 
\begin{eqnarray}
	k^2 &=& \frac{-\beta \pm \sqrt{\beta^2 -4 \alpha \gamma}}{2\alpha} \label{3-47}
\end{eqnarray}
is positive. 
In this case, Eq. (\ref{PTcondition}) is satisfied for a specific value of $\xi_2-\xi_1$,
and the perfect transmission would occur incidentally.
For instance, when we choose the parameters $L_1^{(+)}=3$, $L_1^{(-)}=1$, $L_2^{(+)}=4$, $L_2^{(-)}=-4$, and $\mu_1=\mu_2 = \frac{\pi}{2}$,
Eq. (\ref{3-47}) becomes $k^2=1.25$.
Equation (\ref{PTcondition}) is satisfied when 
\begin{eqnarray}
	\sqrt{1.25} (\xi_2-\xi_1) = -{\rm Arctan} \left( \frac{\sqrt{1.25}}{2} \right) + n\pi \ \ \ {\rm for}\ n=1,2,3, \cdots. \label{3-61}
\end{eqnarray}
The transmission probability in this case is shown as a function of $k$ in Fig.9.  
%
%
\begin{figure}[ht]
\begin{center}
	\includegraphics[width=100mm]{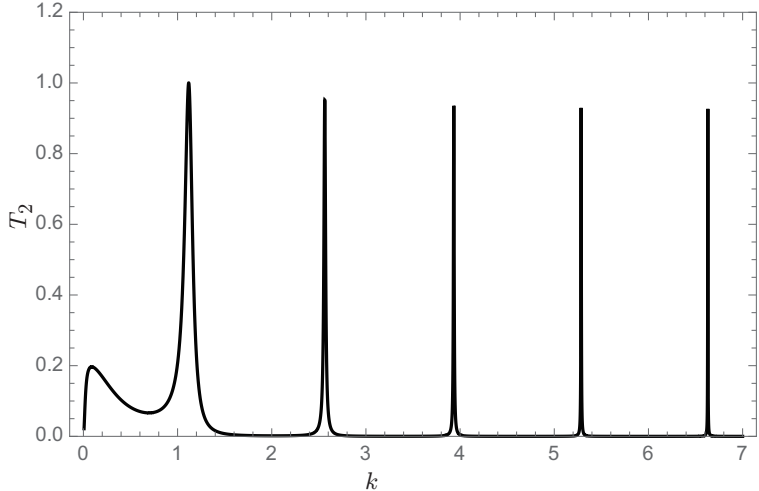}
	\caption{
	The transmission probability $T_2$ is shown as a function of $k$.
	Here we adopt $L_1^{(+)}=3$, $L_1^{(-)}=1$, $L_2^{(+)}=4$, $L_2^{(-)}=-4$, $\mu_1=\mu_2= \frac{\pi}{2}$ and $\xi_2-\xi_1 \simeq 2.354 $ ( $n=1$ in Eq. (\ref{3-61}) ).
	It is shown that perfect transmission occurs ($T_2=1$) at the only $k=\sqrt{1.25}$.
	 }
\end{center}
\end{figure}

\section{Summary}
\label{sec4}

We have investigated the scattering of a quantum particle by two independent point interactions in one dimension.
By considering incident plane wave, we found the condition Eq. (\ref{PT}) under which perfect transmission occurs.
The condition was written as the relation between reflection amplitudes at each point interaction
and is independent of parametrization of the point interactions.
Furthermore, we provided the physical interpretation of the resonance condition.
When the perfect transmission occurs, 
it was shown that each of the transmitting, reflecting, and intermediate plane wave can be expressed as a superposition of an infinite number of the waves with an aligned phase.
The parameter space for two independent point interactions is given by the group $U(2)\times U(2)$, and described by eight parameters $L_{i}^{(+)}$, $L_{i}^{(-)}$, $\mu_{i}$, and $\nu_{i}$ $(i=1,2)$.
Performing an analysis on the whole parameter space, we identified the all parameter region under which perfect transmission occurs.

For the future works, we can consider the 
scattering through $N$ independent multiple point interactions, or through Y-junctions.
It should be noted that the poles of $S$ matrix would also be important for the future works.


 \appendix

 \section{A parametrization of $U(2)$}
 
 In this appendix, we investigate the parametrization of $U(2)$ in Eq. (\ref{U1}). 
An arbitrary $2 \times 2$ unitary matrix $U \in U(2)$ can be  diagonalized by an appropriate unitary matrix, $V \in U(2)$, as
 \begin{eqnarray}
 	 V U V^{\dagger} =  D=  \left( \begin{array}{cc}
		e^{i \theta_+} & 0 \\
		0 & e^{i \theta_-} 
	\end{array}
	\right), \label{A1}
 \end{eqnarray}
 or
 \begin{eqnarray}
	U= V^{\dagger} D V, \label{A2}
\end{eqnarray}
 where 
$e^{i \theta_+}$ and $e^{i \theta_-}$ are eigenvalues of $U$. 
It is well known that the Euler angle representation of $SU(2)$ is
 \begin{eqnarray}
 	 e^{i\frac{\lambda}{2} \sigma_3} e^{i \frac{\mu}{2} \sigma_2} e^{i \frac{\nu}{2} \sigma_3},
\end{eqnarray}
where $\sigma_i\ (i=1,2,3)$ are the Pauli matrices.  
An arbitrary element of $U(2)$ can be expressed as the product of an element of $SU(2)$ and an element of $U(1)$.
Thus, the unitary matrix $V$ which diagonalize $U$ can be written as
 \begin{eqnarray}
 	V= e^{i \eta} e^{i\frac{\lambda}{2} \sigma_3} e^{i \frac{\mu}{2} \sigma_2} e^{i \frac{\nu}{2} \sigma_3}. \label{A4}
\end{eqnarray}
When we substitute Eq. (\ref{A4}) into Eq. (\ref{A2}), $\eta$ and $\lambda$ vanish.
Therefore, $U$ is generally  written as
 \begin{eqnarray}
	U= V^{\dagger} D V,\ \ \ V=  e^{i \frac{\mu}{2} \sigma_2} e^{i \frac{\nu}{2} \sigma_3}.
\end{eqnarray}
The structure of this parameter space was studied in detail in \cite{Mobius}.


\begin{thebibliography}{00}

\bibitem{U21}
	M. Reed, B. Simon, Methods of Modern Mathematical Physics, Vol. II, Academic Press, New York, 1980.
\bibitem{U22}
	P. \v{S}eba, Czech. J. Phys. {\bf 36} (1986) 667.
\bibitem{U23}
	    S. Albeverio, F. Gesztesy, R. H{\o}egh-Krohn, H. Holden,
    Springer, New York, 1988.
\bibitem{duality1}
	T. Cheon, T. Shigehara, Phys. Rev. Lett. {\bf 82} (1999) 2536, quant-ph/9806041.
\bibitem{duality2}
	I. Tsutsui, T. F{\"u}l{\"o}p, T.Cheon, J.Phys. Soc. Jpn. {\bf 69} (2000) 3473, quant-ph/0003069.
\bibitem{susy1}
	T.Uchino, I. Tsutsui, Nucl. Phys. B {\bf 662} (2003) 447, quant-ph/0210084.
\bibitem{susy2}
	T. Uchino, I. Tsutsui, J. Phys. A {\bf 36} (2003) 6493, hep-th/0302089.
\bibitem{susy3}
	T. Nagasawa, M. Sakamoto, K. Takenaga, Phys. Lett. B {\bf 562} (2003) 358, hep-th/0212192.
\bibitem{susy4}
	T. Nagasawa, M. Sakamoto, K. Takenaga, Phys. Lett. B {\bf 583} (2004) 357, hep-th/0311043.
\bibitem{susy5}
	T. Nagasawa, M. Sakamoto, K. Takenaga, J. Phys. A {\bf 38} (2005) 8053, hep-th/0505132.
\bibitem{GP1}
	T. Cheon, Phys. Lett. A {\bf 248} (1998) 285, quant-ph/9803020.
\bibitem{GP2}
	P. Exner, H. Grosse, math-ph/9910029.
\bibitem{GP3}
	S. Ohya, Ann. Phys. {\bf 351} (2014) 900, arXiv:1406.4857 [hep-th].
\bibitem{AH1}
	T. Cheon, T. F{\"u}l{\"o}p, I.Tsutsui, Ann. Phys. {\bf 294} (2001) 1, quant-ph/0008123.	
\bibitem{SC16}
	D.W.L Sprung, H. Wu, J. Matorell, Am. J. Phys. {\bf 61} (1993) 1118.	

\bibitem{SC2}
	P. L. Christiansen, H. C. Arnbak, A. V. Zolotaryuk, V. N. Ermakov, Y. B. Gaididei, J. Phys. A : Math. Gen. {\bf 36} (2003) 7589.
\bibitem{SC3}
	A. V. Zolotaryuk, P. L. Christiansen, S. V. Iermakova, J. Phys. A : Math. Gen. {\bf 39} (2006) 9329.

\bibitem{SC4}
	A. V. Zolotaryuk, P. L. Christiansen, S. V. Iermakova, J. Phys. A : Math. Theor. {\bf 40} (2007) 5443.
\bibitem{SC15}
	F. M. Toyama, Y. Nogami, J. Phys. A : Math. Theor. {\bf 40} (2007) F685.
\bibitem{SC5}
	M. Gadella, J. Negro, L. M. Nieto, Phys. Lett. A {\bf 373} (2009) 1310.
\bibitem{SC6}
	A. V. Zolotaryuk, Phys. Lett. A {\bf 374} (2010) 1636, arXiv:0905.0974 [math-ph].
\bibitem{SC7}
	S. L. S. Ko{\v c}inac, V. Milanovi{\v c}, Mod. Phys. Lett. B {\bf 26} (2012) 1250092.
\bibitem{SC8} 
	A. V. Zolotaryuk, Phys. Rev. A {\bf 87} (2013) 052121, arXiv:1303.4162 [quant-ph].
\bibitem{SC10}
	A. V. Zolotaryuk, Y. Zoloraryuk, Int. J. Mod. Phys. B {\bf 28} (2014) 1350203.
\bibitem{SC11}
	A. V. Zolotaryuk, Y. Zoloraryuk, J. Phys. A : Math. Theor. {\bf 48} (2015) 035302.
\bibitem{SC12}
	A. V. Zolotaryuk, J.Phys. A : Math. Theor. {\bf 48} (2015) 255304.
\bibitem{SC13}
	M. Gadella, J. Mateos-Guilarte, J. M. Munoz-Castaneda, L. M. Nieto, J. Phys. A : Math. Thoer. {\bf 49} (2016) 015204.
\bibitem{SC14}
	M. A. Lee, J. T. Junardi, L. A. Manzoni, E. A. Nyquist, Front. Phys. {\bf 4} (2016) 10.

\bibitem{Mobius}
	I. Tsutsui, T. F{\"u}l{\"o}p, T. Cheon, J. Math. Phys. {\bf 42} (2001) 5687, quant-ph/0105066.
\bibitem{SC1}
	A. G. M. Schmidt, B. K. Cheng, M. G. E. da Luz, Phys. Rev. A {\bf 66} (2002) 062712, quant-ph/0211193.
\bibitem{cheon}
	P. Heja\v{c}{\'i}k, T. Cheon, Phys. Lett. A {\bf 356} (2006) 290, quant-ph/0512239.
\bibitem{KNT}
	K. Konno, T. Nagasawa, R. Takahashi, Ann. Phys. {\bf 375} (2016) 91, arXiv:1605.05418 [quant-ph].



\end{thebibliography}


\end{document}